\documentclass[aps,prd]{revtex4}

\begin{document}

\title{The Two-loop Massless
$\frac{\lambda}{4!}\,\varphi^{4}$ Model\\ in  Non-translational
Invariant Domain}
\author{ M. Aparicio Alcalde$~\!^{(a)}$\thanks{E-mail:aparicio@cbpf.br},
G. Flores Hidalgo$~\!^{(b)}$\thanks{E-mail:gflores@ift.unesp.br}
and N. F. Svaiter$~\!^{(a)}$\thanks{nfuxsvai@cbpf.br}}
\address{$^{(a)}$ Centro Brasileiro de Pesquisas F\'{\i}sicas-CBPF,
Rua Dr. Xavier Sigaud 150, Rio de Janeiro, RJ, 22290-180, Brazil\\
$^{(b)}$ Instituto de F\'{\i}sica Te\'orica-IFT/UNESP, Rua
Pamplona 145, S\~ao Paulo, SP, 01405-900, Brazil}

\begin{abstract}
We study the $\frac{\lambda}{4!}\varphi^{4}$ massless scalar field
theory in a four-dimensional Euclidean space, where all but one of
the coordinates are unbounded. We are considering Dirichlet
boundary conditions in two hyperplanes, breaking the translation
invariance of the system. We show how to implement the
perturbative renormalization up to two-loop level of the theory.
First, analyzing the full two and four-point functions at the
one-loop level, we shown that the bulk counterterms are sufficient
to render the theory finite. Meanwhile, at the two-loop level, we
have to introduce also surface counterterms in the bare lagrangian
in order to make finite the full two and also four-point Schwinger
functions.

\vspace{0.34cm}
 \noindent PACS number(s): 03.70+k, 04.62+v

\end{abstract}

\maketitle

\section{Introduction}

In this paper we are interested to show how to implement the
renormalization procedure up to two-loop level in the massless
$\frac{\lambda}{4!}\varphi^{4}$ scalar field theory, defined in a
four-dimensional Euclidean space with one compactified dimension.
Our aim is to shed light on the renormalization procedure in a
system defined in a domain where translational symmetry is broken,
which must be done for example in the high temperature dimensional
reduced quantum chromodynamics (QCD), defined in a finite region.

Quantum chromodynamics is a non-abelian Yang-Mills theory with
gauge group $SU(3)$. Since it is assumed that the fermions of the
theory transform according to the fundamental representation of
the gauge group, each flavour of quark is a triplet of the color
group $SU(3)$. Gauge bosons transform according to the adjoint
representation. The interaction between the quarks is mediated by
the gluons. Due to the non-abelian structure of the theory, the
gluons couple not only with the quarks but have also cubic and
quartic self-interaction. The self-interaction of the gluons
provides the anti-screening of the color charge in QCD. This is
responsible for asymptotic freedom and presumably confinement.

The confinement-deconfinement phase transition in QCD may occur in
usual matter at sufficiently high temperature or if is strongly
compressed \cite{shuryak} \cite{mc} \cite{gross}. In
ultra-relativistic heavy ion collisions, we expect that the plasma
of quarks and gluons can be produced. We would like to stress
that, although non-equilibrium processes occur in the quark-gluon
plasma in the heavy ion collisions, for simplicity in a first
approximation we can assume a static situation. Just after the
collision hot and compressed nuclear matter is confined in a small
region of the space and in such circumstances the volume and
surface effects become very important.

In the above described physical situation there are two important
points: the first one is that the thermodynamic limit of the
infinite volume system can not be used and therefore finite volume
effects should be investigated and taken into account. The second
point is that the quark-gluon plasma exists in a situation of high
temperature, where using the Matsubara formalism to describe high
temperature QCD, dimensional reduction must occurs \cite{nad}
\cite{jor} \cite{lan} \cite{bra}. Dimensional reduction is based
in the Appelquist-Carrazone  decoupling theorem \cite{ac}. From a
more fundamental theory, the effective Lagrangian density of this
theory can be obtained as some low-energy limit of the fundamental
theory where the heavy modes have been removed. There are some
interesting physical situations where the decoupling theorem can
be used. First, for scalar fields without spontaneous symmetry
breaking. Second, in quantum electrodynamics, where a derivative
expansion of the photon effective action can be obtained by
integrating out the fermionic fields. Also in QCD at least in
lowest order in perturbation theory the decoupling theorem works.
The decoupling theorem is not valid for example in spontaneous
broken gauge theories. It is important to stress that the
non-validity of the decoupling theorem means that low-energy
experiments can provide information about the high energy physics.

Going back to the heavy-ions collision situation, we can assume
that the following scenario appears: in the situation where
dimensional reduction occurs, we have an effective theory for the
gluons field and also finite size effects for these bosonic
fields. To shed light on the renormalization procedure in systems
defined in domain where translational symmetry is broken, as for
example the high temperature dimensional reduced QCD, in this
paper we are interested to investigate scalar models, impose
classical boundary condition over the fields. We hope that this
study will give us some insight over the most interesting and also
more complicated situation as the one mentioned above. Therefore,
in this paper we analyze how to implement the perturbative
renormalization up to two-loop level of
$\frac{\lambda}{4!}\varphi^{4}$ massless scalar field model
defined in a four-dimensional Euclidean space with one
compactified dimension.

Finite size effects and the presence of macroscopic structures in
different field theory models has been extensively studied in the
literature. The critical behavior of the $O(N)$ model in the
presence of a surface was a target of intense investigations
\cite{di}. The same $O(N)$ model was studied in two different
geometries: the periodic cube and the cylinder along one dimension
(the time) and finite and periodic in the $(d-1)$ remaining
dimensions by Brezin and Zinn-Justin \cite{brez}. Finite size
effects in QCD \cite{mit} and also in different field theory
models also has been extensively studied in the literature.
Assuming periodic or antiperiodic boundary conditions for bosonic
and fermionic models respectively, the translation symmetry is
maintained, and surface effects are avoided. Therefore, to avoid
surface effects, quantum fields defined in manifolds with periodic
or anti-periodic boundary conditions in the spatial section was
preferred by many authors \cite{bi}. Nevertheless, the case of
boundaries conditions that break the translational symmetry
deserves our attention.

In the case of hard boundary conditions as for example
Dirichlet-Dirichlet(DD) or Neumann-Neumann (NN), the translational
invariance is lacking. This fact makes the Feynman diagrams harder
to compute that in unbounded space.  Moreover the renormalization
program is implemented in different way from unbounded or
translational invariance systems since some surface divergence
appear \cite{sy}. For translational invariant systems, one can use
the momentum space representation, which is a more convenient
framework to analize the ultraviolet divergences of a theory.
Translational invariance is preserved for momentum conservation
conditions. For non-translational invariant systems a more
convenient representation for the n-point Schwinger functions is a
mixed momentum coordinate space.

Fosco and Svaiter considered the anisotropic scalar model in a
d-dimensional Euclidean space, where all but one of the
coordinates are unbounded. Translational invariance along the
bounded coordinate which lies in the interval $[0,L]$ is broken
because the choice of boundary condition chosen for the
hyperplanes at $z=0$ and $z=L$. Two different possibilities of
boundary conditions was considered: (DD) and also (NN), and the
renormalization of the two-point function  was achieved in the
one-loop approximation \cite{fo}. Further the renormalization of
the four-point function  was achieved in the one-loop
approximation by Caicedo and Svaiter \cite{mario}. Finally Svaiter
\cite{namit} studied the renormalization of the
$\frac{\lambda}{4!}\varphi^{4}$ massless scalar field model in the
one-loop approximation in finite size systems assuming that the
system is in thermal equilibrium with a reservoir. Also, still
studying surface, edge and corners effects, Rodrigues and Svaiter
\cite{ro}  analysed  first the renormalized vacuum fluctuations
associated with a massless real scalar field, confined in the
interior of a rectangular infinitely long waveguide. A closed form
of the analytic continuation of the local zeta function in the
interior of the waveguide was obtained and a detailed study of the
surface and edge divergences was presented. Next, these authors
\cite{ro2} studied the renormalized stress tensor associated with
an electromagnetic field in the interior of a rectangular
infinitely long waveguide.

In this paper we will consider an interacting massless scalar
model, in a four-dimensional Euclidean space, where the first
three coordinates are unbounded and the last one lies in the
interval $[0,L]$. We analyze only $DD$ boundary conditions. First
we present an algebraic expression in coordinate space for the
free propagator which let us to identify the divergences of the
n-point Schwinger functions for the interacting theory. This
algebraic expression agrees with the result obtained by Lukosz
\cite{Greiner}. We would like to stress that instead of assume
hard boundary conditions, some authors assumed soft boundary
conditions and also treated the boundary as a quantum mechanical
object \cite{hard}. Here, we prefer to keep hard classical
boundary conditions.

The organization of the paper is as follows: In section II we
discuss the slab configurations, obtaining some important
expressions for the free propagator in order to understand some
procedures in the divergences identification. In section III the
regularization program is implemented in the one-loop
approximation. In section IV the regularization program is
implemented in the two-loop approximation. Section V contains our
conclusions. In appendix A, expression for the free propagator is
introduced. Throughout this paper we use $\hbar=c=1$.

\section{Classical boundary conditions and some properties of the free propagator}

Let us consider a neutral scalar field with a
$(\lambda\varphi^{4})$ self-interaction, defined in a
$d$-dimensional Minkowski spacetime. The vacuum persistence
functional is the generating functional of all vacuum expectation
value of time-ordered products of the theory. The Euclidean field
theory can be obtained by analytic continuation to imaginary time
allowed by the positive energy condition for the relativistic
field theory. In the Euclidean field theory, we have the Euclidean
counterpart for the vacuum persistence functional, that is, the
generating functional of complete Schwinger functions. The
$(\lambda\varphi^{4})_{d}$ Euclidean theory is defined by these
Euclidean Green's functions. The Euclidean generating functional
$Z(h)$ is formally defined by the following functional integral:
\begin{equation}
Z(h)=\int [d\varphi]\,\, \exp\left(-S_{0}-S_{I}+ \int d^{d}x\,
h(x)\varphi(x)\right), \label{1a}
\end{equation}
where the action that describes a free scalar field is
\begin{equation}
S_{0}(\varphi)=\int d^{d}x\, \left(\frac{1}{2}
(\partial\varphi)^{2}+\frac{1}{2}
m_{0}^{2}\,\varphi^{2}(x)\right), \label{2a}
\end{equation}
and the interacting part, defined by the non-Gaussian
contribution, is
\begin{equation}
S_{I}(\varphi)= \int d^{d}x\,\frac{\lambda}{4!} \,\varphi^{4}(x).
\label{3a}
\end{equation}

In Eq.(\ref{1a}), $[d\varphi]$ is a translational invariant
measure, formally given by $[d\varphi]=\prod_{x} d\varphi(x)$. The
terms $\lambda$ and $m_{0}^{2}$ are respectivelly the bare
coupling constant and mass squared of the model. Finally, $h(x)$
is a smooth function that we introduce to generate the Schwinger
functions of the theory by means of functional derivatives. Note
that we are using the same notation for functionals and functions,
for example $Z(h)$ instead the usual notation $Z[h]$.

In the weak-coupling perturbative expansion, we perform a formal
perturbative expansion with respect to the non-Gaussian terms of
the action. As a consequence of this formal expansion, all the
$n$-point unrenormalized Schwinger functions are expressed in a
powers series of the bare coupling constant $g_{0}$. Let us
summarize how to perform the weak-coupling perturbative expansion
in the $(\lambda\varphi^{4})_{d}$ theory. The Gaussian functional
integral $Z_{0}(h)$ associated with the Euclidean generating
functional $Z(h)$ is
\begin{equation}
Z_{0}(h)=\,{\cal{N}}\int\,[d\varphi]\,\,
\exp\left(-\frac{1}{2}\,\varphi\,K\,\varphi+h\varphi\right).
\label{zz1}
\end{equation}
We are using a compact notation and the first term in the
right-hand side of Eq.(\ref{zz1}) is given by
\begin{equation}
\varphi\,K\,\varphi=\int d^{d}x\,\int d^{d}y\,
\varphi(x)K(m_{0};\,x,y)\varphi(y). \label{zz2}
\end{equation}
The term that couples linearly the field with the external source
is
\begin{equation}
h\varphi=\int d^{d}x\, \varphi(x)h(x). \label{zz3}
\end{equation}

As usual ${\cal{N}}$ is a normalization factor and the symmetric
kernel $K(m_{0};x,y)$ is defined by
\begin{equation}
K(m_{0};x,y)=(-\Delta+m^{2}_{0}\,)\,\delta^{d}(x-y), \label{benar}
\end{equation}
where $\Delta$ denotes the Laplacian in the Euclidean space
$R^{d}$. As usual, the normalization factor is defined using the
condition $Z_{0}(h)|_{h=0}=1$. Therefore
${\cal{N}}=\left(\mbox{det}(-\Delta+m_{0}^{2})\right)^{\frac{1}{2}}$
but, in the following, we are absorbing this normalization factor
in the functional measure. It is convenient to introduce the
inverse kernel, i.e. the free two-point Schwinger function
$G_{0}(m_{0};x-y)$ which satisfies  the identity
\begin{equation}
\int d^{d}z\,G_{0}(m_{0};x-z)K(m_{0};z-y)=\delta^{d}(x-y).
\label{zz4}
\end{equation}
Since Eq.(\ref{zz1}) is a Gaussian functional integral, simple
manipulations, performing only Gaussian integrals, gives
\begin{eqnarray}
\int [d\varphi]\,\, e^{-S_{0}+\int d^{d}x\,
h(x)\varphi(x)}=\exp\left[\frac{1}{2}\int d^{d}x\,
\int
d^{d}y\,\,h(x)\,G_{0}^{(2)}(m_{0};x-y)h(y)\right].
\label{6a}
\end{eqnarray}
Therefore, we have an expression for $Z_{0}(h)$ in terms of the
inverse kernel $G^{(2)}_{0}(m_{0};x-y)$, i.e., in terms of the
free two-point Schwinger function. This construction is
fundamental to perform the weak-coupling perturbative expansion
with the Feynman diagramatic representation of the perturbative
series. The non-Gaussian contribution is a perturbation with
regard to the remaining terms of the action. It is important to
point out that the weak-coupling perturbative expansion can be
defined in arbitrary gemetries, and classical boundary conditions
must be implemented in the two-point Schwinger function. Another
way is to restrict the space of functions that appear in the
functional integral.

We are interested to study finite size systems, where the
translational invariance is broken. In this situation, we are
analyzing the perturbative renormalization for the
$\frac{\lambda}{4!}\varphi^{4}$ massless scalar field model, in
the two-loop approximation. Therefore, let us assume boundary
conditions over the plates for the massless field $\varphi(x)$.
For simplicity we are assuming Dirichlet-Dirichlet boundary
conditions i.e.,
\begin{equation}
\varphi(\vec{r},z)|_{z=0}=\varphi(\vec{r},z)|_{z=L}=0\,,
\label{1}
\end{equation}
for the free field. Since the translational invariance is not
preserved, let us use a Fourier expansion of the fields in the
following form
\begin{equation}
\varphi(\vec{r},z)=\frac{1}{(2\pi)^{\frac{d-1}{2}}} \int
d^{d-1}p\sum_{n}\phi_{n}(\vec{p})\,e^{i\vec{p}.\vec{r}} u_{n}(z),
\label{2}
\end{equation}
where the set $u_{n}(z)$ are the orthonormalized eigenfunctions
associated to the operator $-\frac{d^2}{dz^{2}}$,
$\Bigl(-\frac{d}{dz^{2}}u_{n}(z)=k^{2}_{n}u_{n}(z)\Bigr)$, and
$k_{n}=\frac{n\pi}{L}$, $n=1,2...$. The orthonormal set correspond
to the eigenfunctions of the Hermitian operator
$-\frac{d^2}{dz^{2}}$ defined on a finite interval is given by
\begin{equation}
u_{n}(z)=\sqrt{\frac{2}{L}}\sin(\frac{n\pi z}{L})\,\,\, n=1,2...
\label{7}
\end{equation}
These eigenfunctions satisfy the completeness and orthonormality
relations, i.e.,
\begin{equation}
\sum_{n}u_{n}(z)u_{n}^{*}(z')= \delta(z-z')
\label{3}
\end{equation}
and
\begin{equation}
\int_{0}^{L} dz\, u_{n}(z)u_{n'}^{*}(z)= \delta_{n,n'}.
\label{4}
\end{equation}
Since we are interested to perform the weak coupling expansion,
let us first write the free two-point Schwinger function. This
free two-point Schwinger function can be expressed in the
following form
\begin{equation}
G_{0}^{(2)}(\vec{r},z,z')=\frac{1}{(2\pi)^{d-1}}\int
d^{d-1}p\sum_{n}
e^{i\vec{p}.\vec{r}}u_{n}(z)u_{n}^{*}(z')\,G_{0,n}(\vec{p}),
\label{5}
\end{equation}
where  $G_{0,n}(\vec{p})$ is given by
\begin{equation}
G_{0,n}(\vec{p})=(\vec{p}^{\,2}+k_{n}^{2}+m^{2})^{-1}.
\label{6}
\end{equation}
Next, we will present some properties of the two-point free
Schwinger-function in order to understand the behavior of the
interacting field theory in the presence of macroscopic
structures. Therefore, in order to understand some procedures used
in the identification of the divergences in the Schwinger
functions that will appear in the next section, let us analyze
some properties of the freet two-point Schwinger function.
Substituting Eq.(\ref{7}) and Eq.(\ref{6}) in Eq.(\ref{5}) we get
that the free propagator
$G_{0}^{(2)}(\vec{r}_{1}-\vec{r}_{2},z_{1},z_{2})$ can be written
as
\begin{eqnarray}
G_{0}^{(2)}(\vec{r}_{1}-\vec{r}_{2},z_{1},z_{2})=
 \frac{2}{L} \sum_{n=1}^{\infty}\sin\left(\frac{n\pi
z_{1}}{L}\right)\sin\left(\frac{\,n\pi z_{2}}{L}\right) \int
\frac{d^{d-1}p}{(2\pi) ^{d-1}}
\frac{e^{i\vec{p}.(\vec{r}_{1}-\vec{r}_{2})}}
{\left(\vec{p}^{\,2}+(\frac{n\pi}{L})^{2}+m^{2}\right)}.
\label{free1}
\end{eqnarray}
The next step is to show that the two-point free Schwinger
function can be written in terms of the variables: $r_{12}$,
$z_{12}^-$ and finally $z_{12}^+$, where
$r_{12}=\frac{|\vec{r}_{1}-\vec{r}_{2}|}{L}$,
$z_{12}^-=\frac{z_1-z_2}{L}$ and $z_{12}^+=\frac{z_1+z_2}{L}$
respectively. Working in the four-dimensional case and also in the
massless situation, a straightforward  calculation (see appendix
A) give us that $G_{0}^{(2)}(\vec{r}_{1}-\vec{r}_{2},z_{1},z_{2})$
can be written as:
\begin{eqnarray}
G_{0}^{(2)}(\vec{r}_{1}-\vec{r}_{2},z_{1},z_{2})=
\frac{1}{16\pi^2L^2}
\sum_{k=-\infty}^{\infty}\left\{\frac{1}{(k-\frac{|z_{12}^-|}{2})^2
+(\frac{r_{12}}{2})^2}-\frac{1}{(k-\frac{z_{12}^+}{2})^2+
(\frac{r_{12}}{2})^2}\right\}. \label{free2}
\end{eqnarray}
The former expression for the two-point Schwinger function was
obtained also by Lukosz \cite{Greiner} using the image method.
Performing the summations in Eq.(\ref{free2}) (see appendix A), it
is possible to find a closed expression for
$G_{0}^{(2)}(\vec{r}_{1}-\vec{r}_{2},z_{1},z_{2})$. We get
\begin{eqnarray}
G_{0}^{(2)}(\vec{r}_{1}-\vec{r}_{2},z_{1},z_{2})= \frac{\sinh(\pi
r_{12})}{16\pi L^2~\! r_{12}}\left\{\frac{\sin(\frac{\pi
z_1}{L})\sin(\frac{\pi
 z_2}{L})}{\left[\sinh^2(\frac{\pi r_{12}}{2})+\sin^2(\frac{\pi z_{12}^-}
{2})\right]\left[\sinh^2(\frac{\pi r_{12}}{2})+\sin^2(\frac{\pi
z_{12}^+}{2})\right]}\right\}.
\label{free3}
\end{eqnarray}
It is not difficult to show that the two-point Schwinger function
$G_{0}^{(2)}(\vec{r}_{1}-\vec{r}_{2},z_{1},z_{2})$
satisfies the following properties:
\\
 $(i)$ The free two-point Schwinger function is not negative,
i.e., $G_{0}^{(2)}(\vec{r}_{1}-\vec{r}_{2},z_{1},z_{2})\geq 0\;,$
 for $z_1,\,z_2\in
 [0,L]$ and $\vec{r}_{1},\,\vec{r}_{2}\in {\cal{R}}^3$, since we are
 working in an Euclidean space.
\\
$(ii)$ The free two-point Schwinger function is zero when one of
its points are evaluated on the boundaries
\begin{eqnarray}
G_{0}^{(2)}(\vec{r}_{1}-\vec{r}_{2},0,z_{2})=G_{0}^{(2)}(\vec{r}_{1}-
\vec{r}_{2},L,z_{2})=
G_{0}^{(2)}(\vec{r}_{1}-\vec{r}_{2},z_{1},0)=G_{0}^{(2)}(\vec{r}_{1}-
\vec{r}_{2},z_{1},L)=0\,,
 \nonumber
\end{eqnarray}
since we are assuming Dirichlet boundary conditions.
\\
$(iii)$ The free two-point Schwinger function contain the usual
bulk divergences, i.e., when
$(\vec{r}_{1},z_1)=(\vec{r}_{2},z_2)$, it is singular. From the
Eq.(\ref{free2}) we can identify three singular terms. Splitting
the free two-point Schwinger function in the singular and regular
terms we have:
\begin{eqnarray}
G_{0}^{(2)}(\vec{r}_{1}-\vec{r}_{2},z_{1},z_{2})&=&
\frac{1}{4\pi^2L^2} \left\{\frac{1}{(z_{12}^-)^2+r_{12}^2}-
\frac{1}{(z_{12}^+)^2+r_{12}^2}- \frac{1}{(2-z_{12}^+)^2+r_{12}^2}
\right\}\nonumber\\
&&+\frac{1}{4\pi^2L^2}\left\{ \sum_{k=-\infty \atop k\neq
0}^{\infty}\frac{1}{(2k-|z_{12}^-|)^2 +r_{12}^2} -\sum_{k=-\infty
\atop k\neq
0,1}^{\infty}\frac{1}{(2k-z_{12}^+)^2+r_{12}^2}\right\}\;.
\label{free4}
\end{eqnarray}
The first term of the right side of the last equation, is singular
only when $\vec{r}_{1}=\vec{r}_{2}$ and $z_1=z_2$. This is the
term that carries the usual bulk divergences. The second term is
singular only when $z_1=z_2=0$ and $\vec{r}_{1}=\vec{r}_{2}$. The
third term is singular only when $z_1=z_2=L$ and
$\vec{r}_{1}=\vec{r}_{2}$. These two terms mentioned previously,
carries surface divergences. Finally the two last terms do not
have singularities.
\\
$(iv)$ When $|\vec{r}_{1}-\vec{r}_{2}|/L\gg 1$ the free
propagator behaves like
\begin{eqnarray}
G_{0}^{(2)}(\vec{r}_{1}-\vec{r}_{2},z_{1},z_{2})\sim
\frac{1}{2\pi L^2}\frac{e^{-\pi
r_{12}}}{r_{12}}\sin\left(\frac{\pi z_1}{L}\right)
\sin\left(\frac{\pi z_2}{L}\right)\;,
\end{eqnarray}
which shows an exponential convergence behavior.
\\
$(v)$ The integral of the variable $\{\vec{r},z\}$ on a
neighborhood around $\{\vec{r}\,',z'\}$ of the free propagator is
finite, i.e., $\int_R d^3r\, dz
\;G_{0}^{(2)}(\vec{r}-\vec{r}\,',z,z')<\infty$. See
Fig.(\ref{pri}).

Property $(v)$ allow us to shows that the external legs of the
Feynman diagrams do not create divergences. Let us suppose we have
the integral corresponding to some Feynman diagram,
\begin{eqnarray}
\int_R d^3r dz
\;G_{0}^{(2)}(\vec{r}-\vec{r}\,',z,z')\;F(\vec{r}\,',z')\;,
\label{fey1}
\end{eqnarray}
where $G_{0}^{(2)}(\vec{r}-\vec{r}\,',z,z')$ is some external leg
and $F(\vec{r}\,',z')$ describes the remainder part of the
diagram. Now in order to proceed we have to use the following
statement: for two continuous and positives functions $f(\vec{x})$
and $g(\vec{x})$ defined in a finite region $R$ with the exception
of the point $\vec{x}_1$ where $f(\vec{x})$ diverges, then the
integral $I=\int_{R}d^dx\,f(\vec{x})g(\vec{x})$ is finite, if and
only if $I'=\int_{V}d^dx\,f(\vec{x})$ is finite on some
neighborhoods $V$ of the point $\vec{x}_1$. With the property
$(v)$ and the statement before we can see that external legs from
the Feynman diagrams do not generate divergences.
\begin{figure}[ht]

\begin{picture}(100,100)
\put(0,0){\line(0,1){100}} \put(100,0){\line(0,1){100}}

\put(30,30){\line(1,1){50}} \put(30,30){\circle{20}}

\put(70,85){{\small $(\vec{r},z)$}} \put(15,5){{\small
$(\vec{r}\,',z')$}} \put(20,42){R}
\end{picture}
\caption[region]{}
\label{pri}
\end{figure}
\section{Regularized two and four-point Schwinger functions at
one-loop order}

In this section we identify the divergent contribution in the two
and four-point Schwinger function at one-loop level. Essentially
we use the Eq.(\ref{free4}) in the 1PI diagrams of the Green
functions considering their external legs, and the integrations in
the coordinate space. We write  Eq.(\ref{free4}) as
\begin{eqnarray}
G_{0}^{(2)}(\vec{r}_{1}-\vec{r}_{2},z_{1},z_{2})=
\frac{1}{4\pi^2L^2} \left[\frac{1}{(z_{12}^-)^2+r_{12}^2}-
\frac{1}{(z_{12}^+)^2+r_{12}^2}-
\frac{1}{(2-z_{12}^+)^2+r_{12}^2}+h(r_{12},z_1,z_2)\right]
 \label{free5}
\end{eqnarray}
where $h(r_{12},z_1,z_2)$ is given by
\begin{eqnarray}
h(r_{12},z_1,z_2)=\frac{1}{4}\sum_{k=-\infty \atop k\neq
0}^{\infty}\frac{1}{(k-\frac{|z_{12}^-|}{2})^2
+(\frac{r_{12}}{2})^2}-\frac{1}{4}\sum_{k=-\infty \atop k\neq
0,1}^{\infty}\frac{1}{(k-\frac{z_{12}^+}{2})^2+(\frac{r_{12}}{2})^2}.
\end{eqnarray}
 From the property $(iii)$ we see that the three first
contributions in the right hand side of Eq.(\ref{free5}) have
singularities. Otherwise, the last term is finite in the whole
domain where we defined the model.
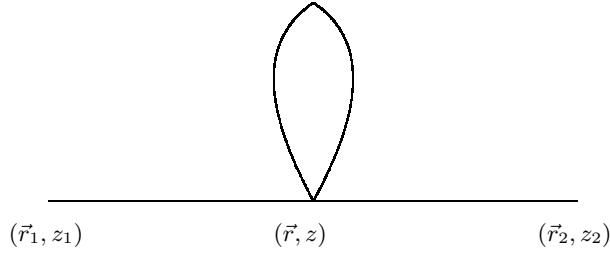
\begin{figure}[ht]
\begin{picture}(100,100)
\put(-40,25){\line(1,0){100}} \put(-55,10){{\small
$(\vec{r}_1,z_1)$}}

\put(60,25){\line(1,0){100}} \put(45,10){{\small $(\vec{r},z)$}}

\put(145,10){{\small $(\vec{r}_2,z_2)$}}
\qbezier(60,25)(30,80)(60,100)
 \qbezier(60,25)(90,80)(60,100)
\end{picture}
\caption[region] {The two-point function at one-loop level}
\label{fig1}
\end{figure}
After this briefly introduction, we are able to study the
interacting theory. Let us start analyzing the tadpole diagram,
displayed in Fig.(\ref{fig1}), from which we can write the
expression for the one loop two-point Schwinger function
$G_{1}^{(2)}(\vec{r}_1-\vec{r}_2,z_1,z_2)$.
We have that
\begin{eqnarray}
G_{1}^{(2)}(\vec{r}_1-\vec{r}_2,z_1,z_2)=\frac{\lambda}{2}\int
d^3r dz \;G_{0}^{(2)}(\vec{r}_1-\vec{r},z_1,z)
\;G_{0}^{(2)}(0,z,z)\;G_{0}^{(2)}(\vec{r}_2-\vec{r},z_2,z)\;.
\label{16}
\end{eqnarray}
In the following we are generalizing the results obtained by Fosco
and Svaiter \cite{fo}.  Let us begin studying the quantity
$G_{0}^{(2)}(0,z,z)$ that appear in the tadpole defined in
Eq.(\ref{16}). From Eq.(\ref{free4}) we get that
$G_{0}^{(2)}(0,z,z)$ can be written as
\begin{eqnarray}
G_{0}^{(2)}(0,z,z)=
\frac{1}{4\pi^2L^2}\left[A-\frac{1}{(2z/L)^2}-\frac{1}{(2-2z/L)^2}
+\sum_{k=-\infty \atop
k\neq0}^{\infty}\frac{1}{(2k)^2}-\sum_{k=-\infty \atop k\neq0,1}^
{\infty}\frac{1}{(2k-2z/L)^2}\right]\;,
 \label{17}
\end{eqnarray}
where $A$ is given by
\begin{eqnarray}
A&=&\lim_{(z_1,\vec{r}_1)\rightarrow(z_2,\vec{r}_2)}\frac{L^2}{(z_1-z_2)^2+
|\vec{r}_1-\vec{r}_2|^2}\nonumber\\
&=&\lim_{\Lambda\rightarrow\infty}\frac{L^2S_4}{8\pi^2}\Lambda^2\;,
\label{19}
\end{eqnarray}
$S_d=\frac{2\pi^{d/2}}{\Gamma(d/2)}$ and $\Lambda$ is an
ultraviolet cutoff. In the
same way, from Eq.(\ref{17}) by performing the summations,  we get for
$G_{0}^{(2)}(0,z,z)$
\begin{eqnarray}
G_{0}^{(2)}(0,z,z)=\frac{1}{4\pi^2L^2}\left[A+\frac{\pi^2}{12}-
\frac{\pi^2}{4}\frac{1}{\sin^2(\pi
z/L)}\right].
 \label{18}
\end{eqnarray}
Substituting Eq.(\ref{19}) in Eq.(\ref{18}) we obtain
\begin{eqnarray}
G_{0}^{(2)}(0,z,z)=\lim_{\Lambda\rightarrow\infty}\frac{S_4}{32\pi^4}
\Lambda^2+\frac{1}{48L^2}-
\frac{1}{16L^2}\frac{1}{\sin^2(\pi z/L)}.
\label{20}
\end{eqnarray}
The first term in Eq.(\ref{20}) is  a bulk divergence.
Substituting Eq.(\ref{20}) in Eq.(\ref{16}) we get
\begin{eqnarray}
G_{1}^{(2)}(\vec{r}_1-\vec{r}_2,z_1,z_2)&=&\lim_
{\Lambda\rightarrow\infty}\frac{\lambda\,S_4}{64\pi^4}
\Lambda^2\int_R d^3r dz \;G_{0}^{(2)}(\vec{r}_1-\vec{r},z_1,z)
\;G_{0}^{(2)}(\vec{r}_2-\vec{r},z_2,z)\nonumber\\
&&+\frac{\lambda}{96L^2}\int_R d^3r dz
\;G_{0}^{(2)}(\vec{r}_1-\vec{r},z_1,z)
\;G_{0}^{(2)}(\vec{r}_2-\vec{r},z_2,z)\nonumber\\
&&-\frac{\lambda}{32L^2}\int_R d^3r dz
\;\frac{G_{0}^{(2)}(\vec{r}_1-\vec{r},z_1,z)\;G_{0}^{(2)}
(\vec{r}_2-\vec{r},z_2,z)}{\sin^2(\pi z/L)}\,.\nonumber\\
\label{21}
\end{eqnarray}
The first term in the right hand side carries a bulk divergence.
The second term is finite. To see this we analyze the integral by
sectors. Therefore we have
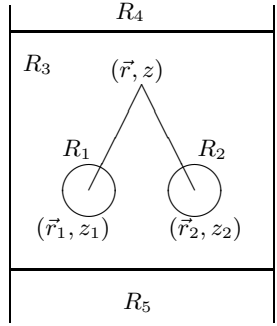
\begin{figure}[ht]

\begin{picture}(130,130)
\put(0,0){\line(0,1){120}} \put(100,0){\line(0,1){120}}

\put(30,50){\line(1,2){20}} \put(30,50){\circle{20}}
\put(20,63){$R_1$} \put(10,33){{\small $(\vec{r}_1,z_1)$}}

\put(70,50){\line(-1,2){20}} \put(70,50){\circle{20}}
\put(71,63){$R_2$} \put(60,33){{\small $(\vec{r}_2,z_2)$}}

\put(38,93){{\small $(\vec{r},z)$}}

\put(0,20){\line(1,0){100}} \put(0,110){\line(1,0){100}}
\put(5,95){$R_3$} \put(40,115){$R_4$} \put(43,5){$R_5$}
\end{picture}
\caption[region]{Regions of integration $R_i$.}
\label{fi3}
\end{figure}
\begin{eqnarray}
\int_R d^3r dz
\;G_{0}^{(2)}(\vec{r}_1-\vec{r},z_1,z)\;G_{0}^{(2)}(\vec{r}_2-\vec{r},z_2,z)
=\int_{R_1}+\int_{R_2}+\int_{R_3}+\int_{R_4}+\int_{R_5}\,,
\label{22}
\end{eqnarray}
where each integral are defined in different regions displayed in
Fig.\ref{fi3}, where the points $(\vec{r}_1,z_{1})$ and
$(\vec{r}_2,z_{2})$ are the centers of the regions $R_1$ and $R_2$
respectively. Using the property $(v)$ we have that the integrals
on $R_1$ and $R_2$ are finite. Since the free propagators
$G_{0}^{(2)}(\vec{r}_1-\vec{r},z_1,z)$ and
$G_{0}^{(2)}(\vec{r}_2-\vec{r},z_2,z) $ presented in Eq.(\ref{22})
does not have divergences on $R_3$ and this region is compact,
then the integral on $R_3$ is finite. The integrals defined in
regions $R_4$ and $R_5$ also are finite since from the property
$(iv)$ the propagator decreases exponentially when one of its
points become far from the other. Thus the integral defined by
Eq.(\ref{22}) is finite. Finally we have to study the third
integral in the right hand side of Eq.(\ref{21}). Note that the
term $\frac{1}{\sin^2(\pi z/L)}$ diverges when $z$ is evaluated on
the boundaries. Nevertheless this integral is convergent, because
the products of $G_{0}^{(2)}(\vec{r}_1-\vec{r},z_1,z)$ and
$G_{0}^{(2)}(\vec{r}_2-\vec{r},z_2,z)$ take away the divergence.
Using Eq.(\ref{free3}), we have that third integral in the right
hand side of Eq.(\ref{21}) is finite. Therefore the one-loop two
point Schwinger function  only has bulk divergence.

Our next step is to analyze the four-point Schwinger function in
the one-loop level. Since the free propagator only has
singularities when its two points are equal or also when this two
points joined are evaluated at the boundaries, we continue our
analysis of the integrals only in the domains where the two
external points of the free propagators take the same values. The
complete four-point function at one-loop level is given by
\begin{eqnarray}
G_{1}^{(4)}(\vec{r}_{1},z_{1},\vec{r}_{2},z_{2},\vec{r}_{3},z_{3},
\vec{r}_{4},z_{4})&=& \frac{\lambda^2}{2}\int d^{d-1}r\int
d^{d-1}r' \int_{0}^{L}dz
\int_{0}^{L}dz'\;G_{0}^{(2)}(\vec{r}_{1}-\vec{r},z_{1},z)
G_{0}^{(2)}(\vec{r}_{2}-\vec{r},z_{2},z)\times
\nonumber\\
&& \left[G_{0}^{(2)}(\vec{r}-\vec{r}\,',z,z')\right]^2\;
G_{0}^{(2)}(\vec{r}_{3}-\vec{r}\,',z_{3},z')\;
G_{0}^{(2)}(\vec{r}_{4}-\vec{r}\,',z_{4},z')\,. \label{four1l}
\end{eqnarray}
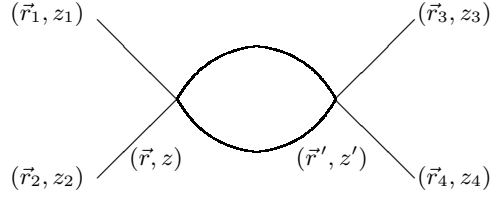
\begin{figure}[ht]

\begin{picture}(100,100)

\put(10,50){\line(-1,1){30}} \put(10,50){\line(-1,-1){30}}
\put(70,50){\line(1,1){30}} \put(70,50){\line(1,-1){30}}
\qbezier(10,50)(20,68)(40,70) \qbezier(10,50)(20,32)(40,30)
\qbezier(40,70)(60,68)(70,50) \qbezier(40,30)(60,32)(70,50)

\put(-8,25){{\small $(\vec{r},z)$}} \put(55,25){{\small
$(\vec{r}\,',z')$}} \put(-53,80){{\small $(\vec{r}_1,z_1)$}}
\put(-53,18){{\small $(\vec{r}_2,z_2)$}} \put(101,80){{\small
$(\vec{r}_3,z_3)$}} \put(101,18){{\small $(\vec{r}_4,z_4)$}}
\end{picture}
\caption[region]{The four point function at one loop.}
\label{fig4a}
\end{figure}
%
\begin{figure}[ht]

\begin{picture}(110,110)
\put(-30,0){\line(0,1){100}} \put(110,0){\line(0,1){100}}

\put(40,50){\circle{50}} \put(20,50){\line(-1,1){20}}
\put(20,50){\line(-1,-1){20}} \put(60,50){\line(1,1){20}}
\put(60,50){\line(1,-1){20}}

\put(19,50){\circle{15}} \put(10,65){$R_1$}

\put(-30,40){\line(1,0){10}} \put(-20,40){\line(0,1){20}}
\put(-30,60){\line(1,0){10}} \put(-27,65){$R_2$}

\put(110,40){\line(-1,0){10}} \put(100,40){\line(0,1){20}}
\put(110,60){\line(-1,0){10}} \put(95,65){$R_3$}
\end{picture}
\caption[region]{Regions of integration for the four point
function} \label{fig5}
\end{figure}
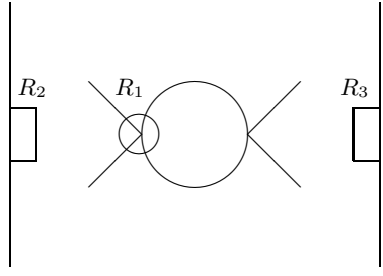
For simplicity, in Fig.(\ref{fig5}) we define three different
regions between the boundaries. The first one, $R_1$ is concerned
when $\{\vec{r}\,',z'\}$ is close to $\{\vec{r},z\}$. In this
region the contribution coming from $\left[G_{0}^{(2)}
(\vec{r}-\vec{r}\,',z,z')\right]^2$ is singular. Nevertheless, we
still have to analyze if this divergent behavior will appear in
the integral defined by Eq.(\ref{four1l}). We will show that the
singularities will appear only as bulk divergences. In the region
$R_2$ ($z, z'\rightarrow 0$ and $\vec{r}\,'\rightarrow\vec{r}$)
the term $\left[G_{0}^{(2)}(\vec{r}-\vec{r}\,',z,z')\right]^2$ is
also divergent. As we will see, this divergent behavior disappears
when we compute the complete four-point function at one-loop
order, defined by Eq.(\ref{four1l}). In the region $R_3$ ($z,
z'\rightarrow L$ and $\vec{r}\,'\rightarrow\vec{r}$) the situation
is identical as in the region $R_2$. Using the same argument that
we used before to analyze the convergence of the integral defined
by Eq.(\ref{fey1}), we can study the convergence of the integral
defined by Eq.(\ref{four1l}) with the amputated external legs.
Therefore we have to study Eq.(\ref{four1l}) with the external
legs amputated. Therefore we have to study the quantity $\int
d^3rdzd^3r'dz'\;
\left[G_{0}^{(2)}(\vec{r}\,'-\vec{r},z',z)\right]^2$. Substituting
Eq.(\ref{free5}) in the former equation we get:
\begin{equation}
\int d^3rdzd^3r'dz'\;
\left[G_{0}^{(2)}(\vec{r}\,'-\vec{r},z',z)\right]^2=
\frac{1}{(4\pi^2L^2)^2} \Bigl(I_1+I_2+I_3+I_4+I_5\Bigr)+\;{\rm
finite~part} \label{nova}
\end{equation}
where the integrals $I_{i}$, $i=1,2,..$ are given by
\begin{eqnarray}
I_1&=&\int
d^3rdzd^3r'dz'\;\frac{1}{\left[(z_{12}^-)^2+r_{12}^2\right]^2}\,,
\\
I_2&=&\int
d^3rdzd^3r'dz'\;\left[\frac{1}{\left[(z_{12}^+)^2+r_{12}^2\right]^2}+
\frac{1}{\left[(2-z_{12}^+)^2+r_{12}^2\right]^2}\right]\,,
\label{36}\\
I_3&=&\int d^3rdzd^3r'dz'\;
\left[-\frac{2}{\left[(z_{12}^-)^2+r_{12}^2\right]
\left[(z_{12}^+)^2+r_{12}^2\right]}
-\frac{2}{\left[(z_{12}^-)^2+r_{12}^2\right]
\left[(2-z_{12}^+)^2+r_{12}^2\right]}\right]\,,
\\
I_4&=&\int
d^3rdzd^3r'dz'\;\frac{2}{\left[(z_{12}^+)^2+r_{12}^2\right]
\left[(2-z_{12}^+)^2+r_{12}^2\right]}\,,
\\
I_5&=&\int d^3rdzd^3r'dz'\;\left[\frac{1}{(z_{12}^-)^2+r_{12}^2}-
\frac{1}{(z_{12}^+)^2+r_{12}^2} -
\frac{1}{(2-z_{12}^+)^2+r_{12}^2}\right]h(r_{12},z_1,z_2)\,.
\label{23}
\end{eqnarray}
Let us investigate each term of Eq.(\ref{nova}). The integral
$I_{1}$ must be analyzed only in the region $R_{1}$. For this
purpose we need an auxiliary result. We can prove that a
continuous and positive function $f(x)$ which does not have
singularities except for $x=0$, and
$M=\int_{-\vec{\epsilon}}^{\vec{\epsilon}}d^dx\,f(w^2)$ where
$w^2=|\vec{w}|^2$, then there exist $\epsilon'$ such that
$M=S_d\int_0^{\epsilon'}dw\,w^{d-1}f(w^2)$ where
$\epsilon<\epsilon'<\sqrt{d}\;\epsilon$. Then we get
\begin{eqnarray}
I_I&=&\int_{R_1}d^3r'dz'\;\frac{1}{[(z_{12}^-)^2+r_{12}^2]^2}
=\int_{\vec{r}-\vec{\epsilon}}^{\vec{r}+\vec{\epsilon}}d^3r'\int_{z-\epsilon}^
{z+\epsilon}dz'\;\frac{1}{[(z-z')^2+|\vec{r}-\vec{r}\,'|^2\,]^2}\nonumber\\
&=&\int_{-\vec{\epsilon}}^{\vec{\epsilon}}\frac{d^4w}{w^4}=S_4\int_{0}^{\epsilon\,'}dw\,\frac{w^3}{w^4}=S_4\ln
w\mid_0^{\epsilon'}.
\end{eqnarray}
Therefore $I_1$ contributes with a bulk divergence of the type as
the one that appears in the theory without boundaries. In the
usual renormalization procedure, the contribution coming from
$I_{1}$ can be  eliminated by the usual counterterms. Concerning
the contribution coming from $I_{2}$ we have that the first term
$1/[(z_{12}^+)^2+r_{12}^2]^2$ is not singular in the region $R_1$.
In the region $R_2$, using the same auxiliary result that we used
before, we can obtain an upper bound to the contribution coming
from this term. We get
\begin{eqnarray}
&&\int_{R_2}d^3rdzd^3r'dz'\;\frac{1}{[(z+z')^2+|\vec{r}-
\vec{r}\,'|^2\,]^2} < \int
d^3r\int_{\vec{r}-\vec{\epsilon}}^{\vec{r}+\vec{\epsilon}}d^3r'
\int_{0}^{\epsilon}\int_{0}
^{\epsilon}dzdz'\;\frac{1}{[z^2+z'\,^2+|\vec{r}-\vec{r}\,'|^2\,]^2}\nonumber\\
&&<\frac{1}{4}\int
d^3r\int_{-\vec{\epsilon}}^{\vec{\epsilon}}d^5w\;\frac{1}{w^2}=
\frac{1}{12}S_5\epsilon'\,^3\int_{R'}d^3r\,.
\end{eqnarray}
Since the region $R\,'\subset R_2$ is finite this integral is
convergent. Next, let us analyze the term
$1/[(2-z_{12}^+)^2+r_{12}^2]^2$ of $I_{2}$ in the region $R_3$.
Since the behavior of the field in each plates (for $z=0$ and
$z=L$) are the same, then the analysis follows the same lines as
previous ones and therefore this contribution is also finite. To
study $I_3$, we consider first the term
$2/[(z_{12}^-)^2+r_{12}^2][(z_{12}^+)^2+r_{12}^2]$. This
expression must be studied in the regions $R_1$ and  $R_2$
respectively. In $R_1$ we can see that the convergence of
\begin{eqnarray}
\int_{R_1}d^3rdzd^3r'dz'\;\frac{1}{[(z-z')^2+|\vec{r}-\vec{r}\,'|^2]
\underbrace{[(z+z')^2+|\vec{r}-\vec{r}\,'|^2]}_{\mbox{finite in}\;
R_1}}\,,
\end{eqnarray}
depend of the convergence of
\begin{eqnarray}
\int_{R_1}d^3rdzd^3r'dz'\;\frac{1}{(z-z')^2+|\vec{r}-\vec{r}\,'|^2}.
\label{nova2}
\end{eqnarray}
From above arguments we have that Eq.(\ref{nova2}) can be written
as
\begin{eqnarray}
\int_{-\vec{\epsilon}}^{\vec{\epsilon}}\frac{d^4w}{w^2}
=S_4\int_{0}^{\epsilon\,'}dw\,w=\frac{S_4\epsilon'\,^2}{2}\,,
\end{eqnarray}
thus Eq.(\ref{nova2}) gives a finite contribution. Now we consider
the first term of $I_3$ in the region $R_2$. For this purpose we
will use the following property. Let us take a continuous and
positive function $f(x)$ which does not have singularities except
for $x=0$, and
$N=\int_{0}^{\vec{\epsilon}}\int_{0}^{\vec{\epsilon}}d^ly\,d^mz\,f(y^2+z^2)$
then there exist $\epsilon'$ in such way
$N=\frac{S_{l+m+2}}{S_{l+1}S_{m+1}}\int_0^{\epsilon'}dw\,w^{l+m+1}f(w^2)$
where $\epsilon'>0$. Using this property, we have for the first
term of Eq.(\ref{36}), in the region $R_2$, that
\begin{eqnarray}
&&\int d^3r\int_0^{\epsilon}dz\int_{z}^{z+
\epsilon}dz'\int_{\vec{r}-\vec{\epsilon}}^
{\vec{r}+\vec{\epsilon}}d^3r'\frac{1}{[(z-z')
^2+|\vec{r}-\vec{r}\,'|^2][(z+z')^2+|\vec{r}-\vec{r}\,'|^2]}\nonumber\\
&&<\int d^3r\int_0^{\epsilon}dz\int_0^{\epsilon}du
\int_{-\vec{\epsilon}}^{\vec{\epsilon}}
d^3v\frac{1}{(u^2+v^2)(z^2+u^2+v^2)}
<\frac{1}{4}S_4\int
d^3r\int_0^{\epsilon}dz\int_0^{\epsilon'}dw\,\frac{w}{(z^2+w^2)}=
\frac{S_3S_4}{2S_1S_2}\epsilon''.
\end{eqnarray}
Therefore the first term of $I_3$ is also finite in $R_2$. The
second term $2/[((z_{12}^-)^2+r_{12}^2)((2-z_{12}^+)^2+r_{12}^2)]$
in $I_3$ must be analyzed also in the regions $R_1$ and $R_3$.
This analysis follows the same lines as the last case, therefore
the contribution coming from this term is also finite.

We have now to study the term $I_4$. Note that
$2/[((z_{12}^+)^2+r_{12}^2)((2-z_{12}^+)^2+r_{12}^2)]$ must be
analyzed in the regions $R_2$ and $R_3$ respectively. Let us start
with the region $R_2$. Using previous arguments we have that the
convergence of $I_4$ depends of the convergence of the following
expression
\begin{eqnarray}
\int_0^{\epsilon}\int_0^{\epsilon}dzdz'\int_{\vec{r}-\vec{\epsilon}}^
{\vec{r}+\vec{\epsilon}}d^3r'\frac{1}{z^2+z'^2+|\vec{r}-\vec{r}\,'|^2}
=\frac{S_5}{4}\int_0^{\epsilon'}dw\,\frac{w^4}{w^2}=\frac{S_5}{12}\epsilon'\,^3\,,
\end{eqnarray}
which is finite. In the region $R_3$ our analysis follow the same
lines as in the region $R_2$, thus the integral in the region
$R_3$ is also finite.

Using the same argument that we used before, is not difficult to
show that the contribution coming from $I_5$ is also finite. We
conclude that the integral given by Eq.(\ref{four1l}) only have
bulk divergences. In this way we can conclude that at one-loop
level the bulk counterterms are sufficient to render the complete
connected Schwinger functions finite. In the next section we will
identify the divergent contribution in the connected two-point
Schwinger functions at the two-loop order.
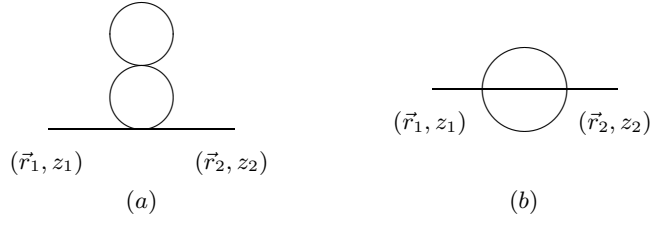
\begin{figure}[ht]

\begin{picture}(100,100)
\put(-75,30){\line(1,0){70}} \put(-90,15){{\small
$(\vec{r}_1,z_1)$}} \put(-20,15){{\small $(\vec{r}_2,z_2)$}}
\put(-40,42){\circle{25}} \put(-40,66){\circle{25}}
\put(-46,0){{\small $(a)$}}

\put(70,45){\line(1,0){70}} \put(55,30){{\small
$(\vec{r}_1,z_1)$}} \put(125,30){{\small $(\vec{r}_2,z_2)$}}
\put(105,45){\circle{30}} \put(99,0){{\small $(b)$}}
\end{picture}
\caption[region] {Two point Schwinger functions at two-loop level}
\label{fig7}
\end{figure}
\section{The divergences in the two-point Schwinger functions at two-loop level}
In this section we will generalize some results obtained by Fosco
and Svaiter \cite{fo} and also by Caicedo and Svaiter
\cite{mario}. We will identify the divergent contribution in the
connected two-point Schwinger functions at the two-loop order. The
diagrams that we are interested to analyze are displayed in
Fig.(\ref{fig7}).
The expression that corresponds to  Fig.(\ref{fig7}-a) is given
by:
\begin{eqnarray}
\frac{\lambda^2}{4}\int d^3r' dz'd^3r dz
\;G_{0}^{(2)}(\vec{r}_1-\vec{r}\,',z_1,z\,')
\left[G_{0}^{(2)}(\vec{r}\,'-\vec{r},z\,',z)\right]^2G_{0}^{(2)}(0,z,z)\,G_{0}^{(2)}
(\vec{r}_2-\vec{r}\,',z_2,z\,'). \label{n5}
\end{eqnarray}
Since the external legs in Eq.(\ref{n5}) does not contribute to
generate divergences, let us consider only the following integral
\begin{eqnarray}
\int d^3r dz
\left[G_{0}^{(2)}(\vec{r}\,'-\vec{r},z\,',z)\right]^2G_{0}^{(2)}(0,z,z).
\label{n6}
\end{eqnarray}
Replacing  Eq.(\ref{20}) in Eq.(\ref{n6}) we get
\begin{eqnarray}
\int d^3r dz
\left[G_{0}^{(2)}(\vec{r}\,'-\vec{r},z\,',z)\right]^2G_{0}^{(2)}(0,z,z)
&=&\lim_{\Lambda\rightarrow\infty}\frac{S_4}{32\pi^4}\Lambda^2
\int
d^3r dz\left[G_{0}^{(2)}(\vec{r}\,'-\vec{r},z\,',z)\right]^2\nonumber\\
& &+\frac{1}{48L^2}\int d^3r dz \left[G_{0}^{(2)}(\vec{r}\,
'-\vec{r},z\,',z)\right]^2\nonumber\\
\nonumber\\
&&-\frac{1}{16L^2}\int d^3r dz \left[G_{0}^{(2)}(\vec{r}\,
'-\vec{r},z\,',z)\right]^2\frac{1}{\sin^2(\pi z/L)}. \label{n7}
\end{eqnarray}
The first term and the second one in Eq.(\ref{n7}) can be
renormalized introducing only bulk counterterms. The most
interesting behavior appears in the last term of this equation.
Note that this unrenormalized quantity contains only bulk
divercences, since the contribution coming from
$\left[G_{0}^{(2)}(\vec{r}\, '-\vec{r},z\,',z)\right]^2$, cancels
the surface divergent behavior generated by the
$\frac{1}{\sin^2(\pi z/L)}$ term. Nevertheless, after the
introduction of a bulk counterterm to render the contribution
$\left[G_{0}^{(2)}(\vec{r}\, '-\vec{r},z\,',z)\right]^2$ finite
between the plates, surface divergences appear. Thus this surface
divergences must be renormalized. After the introduction of
surface and bulk counterterms, the finite contribution coming from
the last term of Eq.(\ref{n7}), up to a finite renormalization
constant, is given by
\begin{eqnarray}
\frac{1}{16L^2}\int d^3r dz
\left[\Bigl(G_{0}^{(2)}(\vec{r}\,'-\vec{r},z\,',z)\Bigr)^2
-\frac{1}{(4\pi^2L^2)^2}\frac{1}{[(z_{12}^-)^2+r_{12}^2]^2}\right]
\left[\frac{1}{\sin^2(\pi z/L)}-\frac{L^2}{(\pi
z)^2}-\frac{L^2}{\pi^2(L-z)^2}\right].
\end{eqnarray}
Therefore this term contains an overlapping between bulk and
surface couterterms.

 We still have to analyze the sunset diagram.
The expression corresponding to the Fig.(\ref{fig7}-b) is given by
\begin{eqnarray}
\frac{\lambda^2}{6}\int d^3r' dz'd^3r dz\,
G_{0}^{(2)}(\vec{r}_1-\vec{r}\,',z_1,z\,')
\left[G_{0}^{(2)}(\vec{r}\,'-\vec{r},z\,',z)\right]^3\;G_{0}^{(2)}
(\vec{r}_2-\vec{r}\,',z_2,z\,'). \label{n8}
\end{eqnarray}
Again, the external legs does not contribute to generate
divergences, and therefore let us study the amputated diagram,
i.e., without external legs. We have
\begin{eqnarray}
\int d^3rdzd^3r'dz'\;
\left[G_{0}^{(2)}(\vec{r}\,'-\vec{r},z',z)\right]^3=\frac{1}{(4\pi^2L^2)^3}
\Bigl(I_1+I_2+...+I_{12}\Bigr)+\;{\rm finite~part}\,, \label{n9}
\end{eqnarray}
where
\begin{eqnarray}
\label{i1}
&&I_1=\int d^3rdzd^3r'dz'\;\frac{1}{[(z_{12}^-)^2+r_{12}^2]^3},\\
\label{i2}
&&I_2=\int d^3rdzd^3r'dz'\;\frac{1}{[(z_{12}^+)^2+r_{12}^2]^3},\\
\label{i3}
&&I_3=\int d^3rdzd^3r'dz'\;\frac{1}{[(2-z_{12}^+)^2+r_{12}^2]^3},\\
\label{i4}
 &&I_4=\int
d^3rdzd^3r'dz'\;\frac{2}{[(z_{12}^-)^2+r_{12}^2]^2[(z_{12}^+)^2+
r_{12}^2]},\\
\label{i5}
 &&I_5=\int
d^3rdzd^3r'dz'\;\frac{2}{[(z_{12}^-)^2+r_{12}^2]
^2[(2-z_{12}^+)^2+r_{12}^2]},\\
\label{i6}
 &&I_6=\int d^3rdzd^3r'dz'\;\frac{2}
{[(z_{12}^-)^2+r_{12}^2][(z_{12}^+)^2+r_{12}^2]^2},\\
\label{i7}
 &&I_7=\int
d^3rdzd^3r'dz'\;\frac{2}{[(z_{12}^-)^2+r_{12}^2][(2-z_{12}^+)^2+
r_{12}^2]^2},\\
\label{i8}
 &&I_8=\int
d^3rdzd^3r'dz'\;\frac{2}{[(z_{12}^+)^2+r_{12}^2]^2
[(2-z_{12}^+)^2+r_{12}^2]},\\
\label{i9}
 &&I_9=\int
d^3rdzd^3r'dz'\;\frac{2}{[(z_{12}^+)^2+r_{12}^2][(2-z_{12}^+)^2+
r_{12}^2]^2},\\
\label{i10}
 &&I_{10}=\int
d^3rdzd^3r'dz'\;\frac{6}{[(z_{12}^-)^2+r_{12}^2]
[(z_{12}^+)^2+r_{12}^2][(2-z_{12}^+)^2+r_{12}^2]},\\
\label{i11}
 &&I_{11}=\int
d^3rdzd^3r'dz'\;\left[\frac{1}{(z_{12}^-)^2+r_{12}^2}-\frac{1}
{(z_{12}^+)^2+r_{12}^2}-\frac{1}{(2-z_{12}^+)^2+r_{12}^2}
\right]^2 \,h(r_{12},z_1,z_2),\\
\label{i12}
 &&I_{12}=\int
d^3rdzd^3r'dz'\;\left[\frac{1}{(z_{12}^-)^2+r_{12}^2}-\frac{1}
{(z_{12}^+)^2+r_{12}^2}-\frac{1}{(2-z_{12}^+)^2+r_{12}^2}\right]\,h^2(r_{12},z_1,z_2).
\end{eqnarray}
Let us analyze each contributions coming from each terms of
Eq.(\ref{n9}). The first integral $I_1$ given by Eq.(\ref{i1}) is
divergent in $R_1$. In general we can show that
\begin{eqnarray}
\int_{R_1} d^3r'dz'\frac{1}{\left[(z_{12}^-)^2+r_{12}^2\right]^n}=
\left\{
\begin{array}{ll}
\mbox{finite}&n<2\\
\infty&n\geq 2
\end{array}
\right.
\end{eqnarray}
Using the above result we can see that the integrals $I_3$, $I_4$
and the first integral of $I_{11}$ are divergent. These integrals
contain bulk divergences which must be removed introducing bulk
counterterms. Next let us analyze the contribution coming from the
integral $I_2$ in the region $R_2$. Using previous arguments and
considering the external legs we get
\begin{equation}
\int_{R_2}d^3rdzd^3r'dz'\;\frac{zz'}{[(z_{12}^+)^2+ r_{12}^2]^3} <
\int d^3r\int_{-\vec{\epsilon}}^{\vec{\epsilon}}\int_
{0}^{\epsilon}\int_{0}^{\epsilon}d^3wdzdz'\;\frac{zz'}{(z^2+z'\,^2+w^2)^3}
<
\frac{S_7}{S_2^2}\epsilon'\int d^3r.
\end{equation}
Therefore this term gives a finite contribution to the
Eq.(\ref{n8}). The contribution from the integral $I_6$ to the
integral must be studied in region $R_2$. In this case we have to
consider the external legs, and the property: let us take a
function $f(x,y)$ positive which does not have singularities
except for $(x,y)=(0,0)$,
$I=\int_0^{\epsilon}\int_0^{\epsilon}dx\,dy\;f(x,y)$ then,
$I<\int_0^{\epsilon}dx\int_x^{x+\epsilon}dy\;f(x,y)+
\int_0^{\epsilon}dy\int_y^{y+\epsilon}dy\;f(x,y)$, we get
\begin{equation}
\int_{R_2}d^3rdzd^3r'dz'\;\frac{zz'}{[(z_{12}^-)^2+
r_{12}^2][(z_{12}^+)^2+r_{12}^2]^2}
<2\int d^3r\int_{-\vec{\epsilon}}^{\vec{\epsilon}}
d^3w\int_{0}^{\epsilon}dz\int_{0}^{\epsilon}
du\;\frac{z(z+u)}{(u^2+w^2)(u^2+z^2+w^2)^2}.
\end{equation}
From above arguments we have that the contribution from the
integral $I_6$ is smaller than
\begin{eqnarray}
&&S_4\int d^3r\int_{0}^{\epsilon}dz\int_{0}^{\epsilon'}
ds\;\frac{z^2s}{(s^2+z^2)^2}
+2S_3\int d^3r\int_{0}^{\epsilon}dz\int_{0}^{\epsilon}
du\int_{0}^{\epsilon'}
ds\;\frac{zus^2}{(u^2+s^2)(u^2+s^2+z^2)^2}\nonumber\\
&&<\frac{S_4S_5}{S_2S_3}\epsilon''\int
d^3r+2\frac{(S_5)^2}{(S_2)^2S_3}\int d^3r\int_{0}^{\epsilon'''}
dw\;\frac{w^4}{w^4}
<\left(\frac{S_4S_5}{S_2S_3}\epsilon''+2\frac{(S_5)^2}{(S_2)^2S_3}
\epsilon'''\right)\int d^3r.
\end{eqnarray}
We conclude that the integral $I_6$ is finite. Also integrating
the contribution coming from the term $I_8$ on $R_2$ we get
\begin{eqnarray}
\int_{R_2}d^3rdzd^3r'dz'\;\frac{1}{[(z_{12}^+)^2+r_{12}^2]
^2\underbrace{[(2-z_{12}^+)^2+r_{12}^2]}_{\mbox{finite}}}.
\end{eqnarray}
Using the fact that the integral
$\int_{R_2}d^3rdzd^3r'dz'\;\frac{1}{[(z_{12}^+)^2+r_{12}^2]^2}$ is
finite, we have that this integral also is convergent in $R_2$.
The contribution from the term $I_{10}$ on $R_2$ is given by
\begin{eqnarray}
\int_{R_2}d^3rdzd^3r'dz'\;\frac{1}{[(z_{12}^-)^2+r_{12}^2]\,
[(z_{12}^+)^2+r_{12}^2]\underbrace{[(2-z_{12}^+)^2+r_{12}^2]}_{\mbox{finite}}}.
\label{n10}
\end{eqnarray}
Since the integral $\int_{R_2}d^3rdzd^3r'dz'
\;\frac{1}{[(z_{12}^-)^2+r_{12}^2]\,[(z_{12}^+)^2+r_{12}^2]}$ is
finite, then the integral defined by Eq.(\ref{n10}) is convergent
in $R_2$. The contribution coming from the terms $I_{11}$ contain
only a bulk divergence. Otherwise, the contributions coming from
the terms $I_{12}$ is finite. We conclude that we need only bulk
counterterms to render the integral defined by Eq.(\ref{n8})
finite. The same analysis can be done for the four-point Schwinger
function in the two-loop approximation. We obtained that only bulk
divergences appear in the full four-point function.

\section{Conclusions}

In this paper we are interested to show how to implement the
renormalization procedure in systems where the translational
invariance is broken by the presence of macroscopic structures.
For the sake of simplicity we are studying the self-interacting
massless scalar field theory in a four-dimensional Euclidean
space. We impose that one coordinate is defined in a compact
domain, introducing two parallel mirrors where we are assuming
Dirichlet-Dirichlet boundaries conditions. Note that although
there are some similarities with the finite temperature field
theory using the Matsubara formalism, in thermal systems appears
only bulk divergences, as for example in the case of system where
we assume periodic boundary conditions. In non-translational
invariant systems, in general to render the theory finite it is
necessary to introduce surface countertems.

In this work we generalize some results obtained by Fosco and
Svaiter \cite{fo} and also by Caicedo and Svaiter \cite{mario}. We
identify the divergences of the Schwinger functions in the
massless self-interacting scalar field theory up to the two-loop
approximation. First, analyzing the full two and four-point
Schwinger functions at the one-loop level, we shown that the bulk
counterterms are sufficient to render the theory finite. Second,
at the two-loop level, we have to introduce surface counterterms
in the bare lagrangian in order to make finite the full two and
also four-point Schwinger functions. The most interesting behavior
appears in the ``double scoop''  diagram given by Eq.(\ref{n5}).
The amputated diagram is given by Eq.(\ref{n6}) and we are
interested in the last term of Eq.(\ref{n7}). This unrenormalized
quantity contains only bulk divercences. Nevertheless, after the
introduction of a bulk counterterm to render the contribution
finite between the plates, surface divergences appear. Thus this
surface divergences must be renormalized. Therefore this term
contains an overlapping between bulk and surface couterterms. This
procedure can be generalized to the n-loop level. The inclusion of
the counterterm in the lagrangian up two-loop level with the full
renormalized action and the general algorithm to identify the
surface and bulk conterterms in the n-loop level will be left to a
future work.

\section{Acknowledgment}
MAA thanks Funda\c{c}\~ao Carlos Chagas Filho de Amparo \`a
Pesquisa do Estado do Rio de Janeiro (FAPERJ) and also thanks
Conselho Nacional do Desenvolvimento Cientifico e Tecnol\'ogico do
Brazil (CNPq), GFH thanks FAPESP, grant 02/09951-3, for full
support and NFS thanks CNPq for partial support.

\newpage

\begin{appendix}
\makeatletter \@addtoreset{equation}{section} \makeatother
\renewcommand{\theequation}{\thesection.\arabic{equation}}

\section{}
In this appendix we will derive an useful representation for the
free two-point Schwinger function. Starting from Eq.(\ref{free1}),
we have that $G_{0}^{(2)}(\vec{r}_{1}-\vec{r}_{2},z_{1},z_{2})$ is
given by
\begin{eqnarray}
G_{0}^{(2)}(\vec{r}_{1}-\vec{r}_{2},z_{1},z_{2})=\frac{2}{(2\pi)^{d-1}L}
\sum_{n=1}^{\infty} \int d^{d-1}p\,\sin\Bigl(\frac{n\pi
z_{1}}{L}\Bigr)\sin\Bigl(\frac{n\pi z_{2}}{L}\Bigr)
\frac{e^{i\vec{p}.(\vec{r}_{1}-\vec{r}_{2})}}
{[\vec{p}^{\,2}+(\frac{n\pi}{L})^{2}+m^{2}]}\,. \label{a1}
\end{eqnarray}
Using the variables $u=\frac{z_1-z_2}{L}$ and
$v=\frac{z_1+z_2}{L}$ defined respectively in the region $u\in
[-1,1]$ and $v\in [0,2]$, and also making use of a trigonometric
identity and performing the sum that appears in Eq.(\ref{a1}) we
obtain \cite{grads}
\begin{eqnarray}
G_{0}^{(2)}(\vec{r}_{1}-\vec{r}_{2},z_{1},z_{2})=\frac{1}{2}
\int\frac{d^{d-1}p}{(2\pi)^{d-1}}
\,\frac{e^{i\vec{p}.(\vec{r}_{1}-
\vec{r}_{2})}}{(\vec{p}^{\,2}+m^{2})^{\frac{1}{2}}}
\left[\frac{\cosh\left(L(1-|u|)(\vec{p}^{\,2}+m^{2})^{\frac{1}{2}}\right)}
{\sinh \left(L(\vec{p}^{\,2}+m^{2})^{\frac{1}{2}}\right)}
-\frac{\cosh\left(L(1-v)(\vec{p}^{\,2}+m^{2})^{\frac{1}{2}}\right)}
{\sinh
\left(L(\vec{p}^{\,2}+m^{2})^{\frac{1}{2}}\right)}\right].\nonumber\\
\end{eqnarray}
Taking $m=0$, $d=4$, and integrating the angular part, it is
possible to show that
$G_{0}^{(2)}(\vec{r}_{1}-\vec{r}_{2},z_{1},z_{2})$ can be written
as
\begin{equation}
G_{0}^{(2)}(\vec{r}_{1}-\vec{r}_{2},z_{1},z_{2})=\frac{-i}{2(2\pi)^2r'L^2}
\int_0^{\infty} dx \left(e^{ixr'}-e^{-ixr'}\right)
\left[\frac{\cosh\left((1-|u|)\,x\right)} {\sinh x}
-\frac{\cosh\left((1-v)\,x\right)}{\sinh x}\right],
\end{equation}
where the variable $r'$ is defined by: $r'\equiv
\frac{|\vec{r}_{1}-\vec{r}_{2}|}{L}$. Making use of the following
integral representation of the product between the Gamma function
and the Riemann zeta function \cite{grads}
\begin{equation}
\int_0^{\infty}dx\; \frac{x^{z-1}e^{-\beta
x}}{e^{px}-1}=\frac{\Gamma(z)}{p^{z}}\zeta(z,\frac{\beta}{p}+1),
\end{equation}
where $Re(z)>1$, $Re(\frac{\beta}{p})>-1$ and the Riemann zeta
function $\zeta(z,q)$ is defined by
\begin{equation}
\zeta(z,q)= \sum_{k=0}^{\infty}\frac{1}{(k+q)^{z}},
\,\,\,\,q\neq\,0,-1,-2..
\end{equation}
then, it is possible to write
$G_{0}^{(2)}(\vec{r}_{1}-\vec{r}_{2},z_{1},z_{2})$ as
\begin{equation}
G_{0}^{(2)}(\vec{r}_{1}-\vec{r}_{2},z_{1},z_{2})=
\frac{1}{16\pi^2L^2}\left[\,\sum_{k=-\infty}^{\infty}\frac{1}
{(k-\frac{|u|}{2})^2 +(\frac{r'}{2})^2}-\sum_{k=-\infty}^{\infty}
\frac{1}{(k-\frac{v}{2})^2+(\frac{r'}{2})^2} \right]. \label{feq}
\end{equation}
Finally, using the following identity:
\begin{equation}
\sum_{k=-\infty}^{\infty}\frac{1}{(k-z)^2+r^2}=
\frac{\pi}{2r}\frac{\sinh(2\pi r)}{\sinh^2(\pi r)+\sin^2(\pi z)},
\end{equation}
we obtain the expression for the two-point Schwinger function that
we need to proceed in our analysis.  Using the above equation in
Eq. (\ref{feq}) we get
\begin{eqnarray}
G_{0}^{(2)}(\vec{r}_{1}-\vec{r}_{2},z_{1},z_{2})=\frac{\sinh(\pi
 r\,')}{16\pi
L^2 r\,'}\left[\frac{\sin(\frac{\pi z_1}{L})\sin(\frac{\pi
z_2}{L})} {\left[\sinh^2(\frac{\pi r\,'}{2})+\sin^2(\frac{\pi u}
{2})\right]\left[\sinh^2(\frac{\pi r\,'}{2})+\sin^2(\frac{\pi
v}{2})\right]}\right].
\end{eqnarray}
\end{appendix}


\newpage

\begin{thebibliography}{50}

\bibitem{shuryak} E. V. Shuryak, Phys. Rep. {\bf 61}, 71 (1980).
\bibitem{mc} L. McLerran, Rev. Mod. Phys. {\bf 58}, 1021 (1986).
\bibitem{gross} D. Gross. R. Pisarski and L. Yaffe, Rev. Mod. Phys. {\bf 53}, 43 (1981).
\bibitem{nad} S. Nadkarni, Phys. Rev. {\bf D27}, 917 (1983),
ibid Phys. Rev. {\bf D38}, 3287 (1988).
\bibitem{jor} A. N. Jourjine, Ann. Phys.{\bf 155}, 305 (1984).
\bibitem{lan} N. D. Landsman, Nuc. Phys. {\bf B322}, 3287 (1988).
\bibitem{bra} E. Braaten and A. Nieto, Phys. Rev. {\bf D55}, 3421 (1996),
A. Nieto, Int. Jour. Mod. Phys. {\bf A12}, 1431 (1997).
\bibitem{ac} T. A. Appelquist and J. Carrazone, Phys. Rev. {\bf D11}, 2856
(1975).
\bibitem{di} H. W. Diehl and S. Dietrich, Z. Phys. {\bf B42}, 65 (1981),
H. W. Diehl, S. Dietrich and E. Eiseriegler, Phys. Rev. {\bf B27},
2937 (1983), A. N. Nemirovsky and K. F. Freed, J. Phys. {\bf A18},
L319 (1985), A. M. Nemirovsky and K. F. Freed, Nucl. Phys. {\bf
B270}, 423 (1986), M. Huhn and V. Dohm, Phys. Rev. Lett. {\bf 61},
1368 (1988), M. Krech and S. Dietrich, Phys. Rev. Lett. {\bf 66},
345 (1991), M. Krech and S. Dietrich, Phys. Rev. {\bf A46}, 1886
(1992).
\bibitem{brez} E. Brezin and J. Zinn-Justin, Nucl. Phys. {\bf B257}, 867 (1985).
\bibitem{mit} A. Chodos, R. L. Jaffe, C. B. Thorn and V. F. Weisskopf,
Phys. Rev. {\bf D9}, 3471 (1974), T. H. Hansson and K. Johnson and
C. Peterson, Phys. Rev. {\bf D26}, 915 (1982); T. H. Hansson and
R. L. Jaffe, Phys. Rev. {\bf D28}, 882 (1983), T. H. Hansson and
R. L. Jaffe, Ann. Phys. {\bf 151}, 204 (1983).
\bibitem{bi} N. D. Birrell and L. H. Ford, Phys. Rev. {\bf D22}, 330 (1980),
L. H. Ford, Phys. Rev. {\bf D21}, 933 (1980); D. J. Toms, Phys.
Rev. {\bf D21}, 928 (1980); D. J. Toms, Ann. Phys. {\bf 129}, 334
(1980); L. H. Ford and N. F. Svaiter, Phys. Rev. {\bf D51}, 6981
(1995).
\bibitem{sy} K. Symanzik, Nucl. Phys. {\bf B190}, 1 (1980).
\bibitem{fo} C. D. Fosco and N. F. Svaiter, J. Math. Phys. {\bf 42},
5185 (2001).
\bibitem{mario} M. Caicedo and N. F. Svaiter, J. Math. Phys. {\bf 45},
179 (2004).
\bibitem{namit} N. F. Svaiter, J. Math. Phys. {\bf 45},
4524 (2004).
\bibitem{ro} R. B. Rodrigues and N. F. Svaiter, Physica {\bf A328}, 466
(2003).
\bibitem{ro2} R. B. Rodrigues and N. F. Svaiter, Physica {\bf A342}, 529
(2004).
\bibitem{Greiner} W. Lukosz, Z. Phys. {\bf 258}, 99 (1973).
\bibitem{hard} A. A. Actor and I. Bender, Phys. Rev. {\bf D52}, 3581 (1995);
F. Caruso, R. De Paola and N. F. Svaiter, Int. Journ. Mod. Phys.
{\bf A14}, 2077 (1999), L. H. Ford and N. F. Svaiter, Phys. Rev.
{\bf D58}, 065007-1 (1998).
\bibitem{grads} I. S. Gradshteyn and I. M. Ryzhik, {\em{Tables of
Integrals, Series and Products }}, Academic Press Inc., New York (1980)

\end{thebibliography}
\end{document}